\documentclass[twoside]{article}

\usepackage{PRIMEarxiv}

\usepackage[utf8]{inputenc} % allow utf-8 input
\usepackage[T1]{fontenc}    % use 8-bit T1 fonts
\usepackage{hyperref}       % hyperlinks
\usepackage{url}            % simple URL typesetting
\usepackage{booktabs}       % professional-quality tables
\usepackage{amsfonts}       % blackboard math symbols
\usepackage{nicefrac}       % compact symbols for 1/2, etc.
\usepackage{microtype}      % microtypography
\usepackage{fancyhdr}       % header
\usepackage{graphicx}       % graphics
\graphicspath{{figs/}}      % organize your images and other figures under figs/ folder

\usepackage{amsmath, amssymb, amsthm}
\usepackage[svgnames]{xcolor}
\usepackage{natbib}
\usepackage{listings}

\title{ipd: An R Package for Conducting Inference on Predicted Data
\thanks{\textit{\underline{Citation}}: 
\textbf{Authors. Title. Pages.... DOI:000000/11111.}} 
}

\author{
  Stephen Salerno \\
  Public Health Sciences, Biostatistics \\
  Fred Hutchinson Cancer Center \\
  Seattle, WA \\
  \texttt{ssalerno@fredhutch.org} \\
  \And
  Jiacheng Miao \\
  Biostatistics and Medical Informatics \\
  University of Wisconsin-Madison \\
  Madison, WI \\
  \texttt{jmiao24@wisc.edu} \\
  \And
  Awan Afiaz \\
  Biostatistics \\
  University of Washington \\
  Seattle, WA \\
  \texttt{aafiaz@uw.edu} \\
  \AND
  Kentaro Hoffman \\
  Statistics \\
  University of Washington \\
  Seattle, WA \\
  \texttt{khoffm3@uw.edu} \\
  \And
  Anna Neufeld \\
  Mathematics and Statistics \\
  Williams College \\
  Williamstown, MA \\
  \texttt{acn2@williams.edu} \\
  \AND
  Qiongshi Lu \\
  Biostatistics and Medical Informatics \\
  University of Wisconsin-Madison \\
  Madison, WI \\
  \texttt{qlu@biostat.wisc.edu} \\
  \And
  Tyler H.~ McCormick \\
  Statistics \\
  Sociology \\
  University of Washington \\
  Seattle, WA \\
  \texttt{tylermc@uw.edu} \\
  \And
  Jeffrey T.~Leek \\
  Public Health Sciences, Biostatistics \\
  Fred Hutchinson Cancer Center \\
  Biostatistics \\
  University of Washington \\
  Seattle, WA \\
  \texttt{jtleek@fredhutch.org} \\
}

\begin{document}

%--- FRONT MATTER ----------------------------------------------

\maketitle

\begin{abstract}
  {\bf Summary:} {\tt ipd} is an open-source {\tt R} software package for the downstream modeling of an outcome and its associated features where a potentially sizable portion of the outcome data has been imputed by an artificial intelligence or machine learning (AI/ML) prediction algorithm. The package implements several recent proposed methods for inference on predicted data (IPD) with a single, user-friendly wrapper function, {\tt ipd}. The package also provides custom {\tt print}, {\tt summary}, {\tt tidy}, {\tt glance}, and {\tt augment} methods to facilitate easy model inspection. This document introduces the {\tt ipd} software package and provides a demonstration of its basic usage.\\
  {\bf Availability:} {\tt ipd} is freely available on CRAN or as a developer version at our GitHub page: github.com/ipd-tools/ipd. Full documentation, including detailed instructions and a usage `vignette' are available at github.com/ipd-tools/ipd.\\
  {\bf Contact:} \href{jtleek@fredhutch.org}{jtleek@fredhutch.org} and \href{tylermc@uw.edu}{tylermc@uw.edu}
\end{abstract}


%--- BODY ------------------------------------------------------

\section{Introduction}
\label{sec:1}

With the rapid advancement of artificial intelligence and machine learning (AI/ML) algorithms, and owing to financial and domain-specific constraints, researchers from a wide range of disciplines increasingly use predictions from pre-trained algorithms as outcome variables in statistical analyses \citep{hoffman2024we}. However, reifying algorithmically-derived values as measured outcomes may lead to potentially biased estimates and anti-conservative inference \citep[e.g., see][]{wang2020methods}. In particular, the statistical challenges encountered when drawing {\it inference on predicted data} (IPD) include: (1) understanding the relationship between the predicted outcomes and their true, unobserved counterparts, (2) quantifying the robustness of the AI/ML models to resampling or uncertainty about the data they were trained on, and (3) appropriately propagating both bias and uncertainty from upstream predictive model into downstream inferential procedures. We refer to methods developed to address these challenges as methods for conducting IPD.

Several recent methods have been proposed for conducting IPD. These include {\it post-prediction inference (PostPI)} by \citet{wang2020methods}, {\it prediction-powered inference (PPI)} and {\it PPI++} by \citet{angelopoulos2023prediction, angelopoulos2023ppi++}, and {\it post-prediction adaptive inference (PSPA)}, as well as PSPA's extensions, {\it POP-TOOLS} and {\it post-prediction sumstats-based inference (PSPS)} by \citet{miao2023assumption, miao2024valid} and \citet{miao2024task}, respectively, {\it prediction-powered bootstrap (PPBoot)} by \citet{zrnic2024note}, and semi-supervised methods, {\it cross-prediction-powered inference (Cross-PPI)} by \citet{zrnic2024cross} and {\it design-based supervised learning (DSL)} by \citet{egami2024using}. Broadly, these methods employ one of two strategies for IPD correction. They either (1) construct pseudo-outcomes designed to resemble the true, unobserved outcome and correct inference on the outcome space, or (2) calibrate the parameter estimates and standard errors directly.

These methods have been developed in quick succession in response to the ever-growing practice of using predicted data directly to conduct statistical inference. To enable researchers and practitioners interested in fully utilizing these state-of-the-art methods, we have developed {\tt ipd}, a comprehensive and open-source software package which implements these existing methods under the umbrella of IPD. Moreover, we provide a convenient wrapper and helper functions so these methods can be compared in numerous inferential settings. This note provides an overview of the package, including installation instructions, basic usage examples, and further documentation. The examples presented here show how to generate data, fit models, and use custom methods provided by the package.

\begin{lstlisting}[backgroundcolor = \color{lightgray!25}, language = R, xleftmargin = 0.25cm, framexleftmargin = 1em]
#-- Install the ipd package from GitHub and load
devtools::install_github("ipd-tools/ipd")
library(ipd)
\end{lstlisting}

\section{Installation and Usage}
\label{sec:2}

{\tt ipd} is implemented in the {\tt R} statistical computing language \citep{rsoftware}. Full documentation, including detailed downloading and installation instructions and usage `vignettes' are available on the package website: github.com/ipd-tools/ipd. The {\tt ipd} package has been successfully tested and runs on all the latest versions of Windows, Mac OS X, Ubuntu (Linux) operating systems. To install the development version of {\tt ipd} from GitHub, one can use the {\tt devtools} package.

We provide a simple example to demonstrate the basic use of this function. Following the notation of \citet{miao2023assumption}, we assume the user has a dataset, $\mathcal{D} = \mathcal{L} \cup \mathcal{U}$, consisting of $\mathcal{L} = \{(Y_i, f_i, \boldsymbol{X}_i, R_i);\ i = 1, \ldots, n\}$ {\it labeled} samples of the observed outcome, $Y$, predicted outcome, $f$, and features of interest, $\boldsymbol{X}$, and $\mathcal{U} = \{(f_i, \boldsymbol{X}_i, R_i);\ i = n + 1, \ldots, n + N\}$ {\it unlabeled} samples, where the outcome is not observed. $R$ is a variable indicating whether the $i$th observation is {\it labeled}, where $R_i = 1$ if the observation is {\it labeled} and $R_i = 0$ otherwise. 

The main function, {\tt ipd}, gives access to the various methods for conducting IPD with different potential estimands. The user supplies a {\tt formula} of the form {\tt Y - f = X1 + X2 + ...}, where {\tt Y} is replaced by the name of the observed outcome variable in the data, {\tt f} is replaced by the name of the predicted outcome, and {\tt X1}, {\tt X2}, {\tt ...}, are the names of the independent variables (features) of interest. The user has the option to supply either the stacked dataset, including the name of the column corresponding to the label indicator, $R$, or the labeled and unlabeled data sets separately, along with the method of choice, and then intended target of inference (estimand). Options for the available methods include {\tt "postpi"} \citep{wang2020methods}, {\tt "ppi"} \citep{angelopoulos2023prediction}, {\tt "ppi\_plusplus"} \citep{angelopoulos2023ppi++}, and {\tt "pspa"} \citep{miao2023assumption}. Current estimands include the population mean ({\tt "mean"}) or $q$th quantile ({\tt "quantile"}, with additional argument {\tt q = q}) of the outcome, linear ({\tt "ols"}), and logistic ({\tt "logistic"}) regression. Future development will include a broader class of exponential family model parameters including multiclass logistic regression, time-to-event models, and causal targets such as average treatment effects (ATE). Additional (optional) arguments include the significance level, {\tt alpha}, for the $100(1 - \alpha)\%$ confidence intervals, and other method-specific arguments.

The function outputs a `glm'-style list with the parameter estimates, standard errors, and confidence limits, as well as additional information about the function call and intermediate estimated quantities (e.g., the estimated relationship model of \citet{wang2020methods} or the estimated tuning parameters of \citet{angelopoulos2023prediction, angelopoulos2023ppi++} and \citet{miao2023assumption}. 

All method-specific and helper functions are documented and exported by the {\tt ipd} package for additional user flexibility. These include a function to generate simulated data, {\tt simdat}, to facilitate exploration of the methods in the absence of real data, as well as {\tt print}, {\tt summary}, {\tt tidy}, {\tt glance}, and {\tt augment} methods to facilitate easy model inspection \citep{tidyr}. In the next section, we provide an illustrative example using simulated data generated by {\tt simdat} for linear regression.

\section{An Example Analysis}
\label{sec:3}

We simulate a continuous outcome for linear regression, with $Y = \beta_1 X_1 + \frac{1}{2} X_2^2 + \frac{1}{3} X_3^3 + \frac{1}{4} X_4^2 + \varepsilon,$ where $X_1$, $X_2$, $X_3$, and $X_4\sim \mathcal{N}(0,1)$, $\beta_1 = 1$, corresponding to the true value of the linear regression coefficient for $X_1$ (our target of inference), and $\varepsilon\sim \mathcal{N}(0, \sigma^2_Y); \sigma_Y = 4$ using the {\tt simdat} function. We generate a stacked dataset of 100 training, 100 labeled, and 1,000 unlabeled observations. We generate predicted outcomes by training a generalized additive model on the training set and making predictions for the labeled and unlabeled sets.

\begin{lstlisting}[backgroundcolor = \color{lightgray!25}, language = R, xleftmargin = 0.25cm, framexleftmargin = 1em]
#-- Generate example data for linear regression
dat_ols <- simdat(n = c(100, 100, 1000), effect = 1, 
    sigma_Y = 4, model = "ols")
\end{lstlisting}

For each method, we calculate the point estimate and corresponding $100(1 - \alpha)\%$ confidence interval for $\beta_1$, where $\alpha = 0.05$ and we have a two-sided hypothesis test. We compare the methods in the package to three additional models, which serve as performance benchmarks: the `oracle' regression, which fits the ideal model on the true, unknown outcome for the unlabeled observations (possible on simulated data), the `naive' regression, which treats the predicted outcomes as if they were the true, unobserved outcomes, and the `classic' regression, which utilizes only the labeled subset of the data:

\begin{lstlisting}[backgroundcolor = \color{lightgray!25}, language = R, xleftmargin = 0.25cm, framexleftmargin = 1em]
dat_ols_l <- dat_ols[dat_ols$set == "labeled",]
dat_ols_u <- dat_ols[dat_ols$set == "unlabeled",]
#-- Benchmark Regressions
fit0 <- lm(Y ~ X1, data = dat_ols_u)  #- Oracle
fit1 <- lm(f ~ X1, data = dat_ols_u)  #- Naive
fit2 <- lm(Y ~ X1, data = dat_ols_l)  #- Classic
#-- PostPI
fit3 <- ipd(Y - f ~ X1, method = "postpi_boot", model = "ols", data = dat_ols, label = "set", nboot = 200)
#-- PPI
fit4 <- ipd(Y - f ~ X1,  method = "ppi", model = "ols", data = dat_ols, label = "set")
#-- PPI++
fit5 <- ipd(Y - f ~ X1, method = "ppi_plusplus", model = "ols", data = dat_ols, label = "set")
#-- PSPA
fit6 <- ipd(Y - f ~ X1, method = "pspa", model = "ols", data = dat_ols, label = "set")
\end{lstlisting}

\begin{figure*}[!ht]%
    \centering
    \caption{Point estimate and corresponding 95\% confidence intervals for four available IPD methods ({\tt postpi}, {\tt ppi}, {\tt ppi\_plusplus} and {\tt pspa}; second row), as compared to three benchmark regressions (oracle, naive, and classical; first row) on 1,000 simulated linear regression datasets.}
    \includegraphics[width=\textwidth]{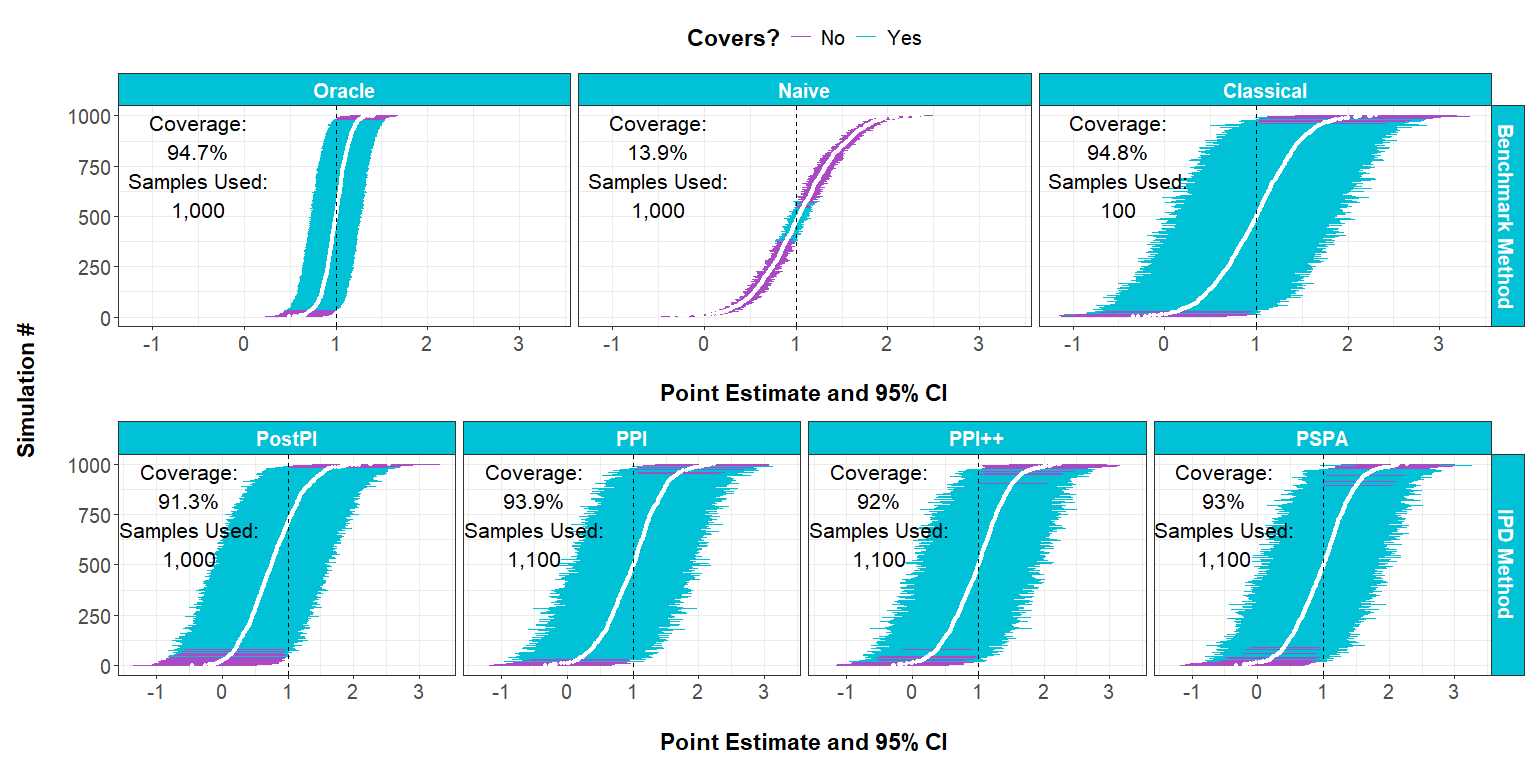}
    \label{fig:results}
\end{figure*}

As a benchmark, the hypothetical `oracle' regression would have the correct estimate and correctly-sized confidence interval if the outcome was measured for each observation, while in practice, the `naive' regression may have a biased point estimate and a confidence interval that is too narrow. The `classical' regression will have the correct estimate but wider confidence interval, as it only uses the labeled subset of the data. The implemented IPD methods all correctly estimate the coefficient and have confidence intervals that are wider than the `oracle' but narrower than the `classic' regression (Figure \ref{fig:results}).

\section{Printing, Summarizing, and Tidying}

The package also provides custom {\tt print}, {\tt summary}, {\tt tidy}, {\tt glance}, and {\tt augment} methods to facilitate easy model inspection. Namely, the {\tt print} method gives an abbreviated summary of the output from the {\tt ipd} function, the {\tt summary} method gives more detailed information about the estimated coefficients, standard errors, and confidence limits, The {\tt tidy} method organizes the model coefficients into a `tidy' format \citep{tidyr}, the {\tt glance} method returns a one-row summary of the model fit, and the {\tt augment} method adds model predictions and residuals to the original dataset. For a more detailed look into using the {\tt ipd} wrapper function and the method-specific individual functions, please refer to the vignettes provided with the package.

\begin{lstlisting}[backgroundcolor = \color{lightgray!25}, language = R, xleftmargin = 0.25cm, framexleftmargin = 1em]
print(fit3)                            #- Print
summ <- summary(fit3)                  #- Summary
print(summ)                            #- Print Summary
tidy(fit3)                             #- Tidy
glance(fit3)                           #- Glance
augmented_df <- augment(fit3)          #- Augment
\end{lstlisting}

\section{Conclusion}

In this note, we present {\tt ipd}, a comprehensive {\tt R} package which implements various recent methods for conducting inference on predicted data. We highlight the usability of this software for practitioners to draw valid inference on algorithmically derived outcomes, as well as the ability to facilitate comparisons for those wishing to further develop methods in the space of IPD. It is our hope that we, and others members of the research community, will maintain and grow this package as the field of IPD continues to mature in the current AI/ML era.

\newpage

%--- FUNDING ---------------------------------------------------

\section*{Acknowledgments}

This work was supported in part by NIH/NIGMS R35 GM144128 and the Fred Hutchinson Cancer Center J.~Orin Edson Foundation Endowed Chair (SS, JTL), NIH/NHGRI U01 HG012039 (JM, QL), NIH/NIMH DP2 MH122405, R01 HD107015, and P2C HD042828 (THM).

%--- REFERENCES ------------------------------------------------

\bibliographystyle{abbrvnat}  
\bibliography{reference}  

%=== END =======================================================

\end{document}